\documentclass{icrc}

\usepackage{times}
\usepackage{graphicx} 

\begin{document}

\title{Estimation of Shower Parameters in Wavefront Sampling Technique}
\author[1]{V. R. Chitnis}
\author[1]{P. N. Bhat}
\affil[1]{Tata Institute of Fundamental Research, Homi Bhabha Road,
Mumbai 400 005, India}

\correspondence{V. R. Chitnis (vchitnis@tifr.res.in)}

\runninghead{Chitnis and Bhat: Estimation of shower parameters }
\firstpage{1}
\pubyear{2001}


\maketitle

\begin{abstract}
Wavefront sampling experiments record arrival times of \v Cerenkov
photons with high precision at various locations in \v Cerenkov pool using
a distributed array of telescopes. It was shown earlier that this photon
front can be fitted with a spherical surface traveling at a speed of light
and originating from a single point on the shower axis. Radius of curvature
of the spherical shower front ($R$) is approximately equal to the height of
shower maximum from observation level. For a given primary species, it is also
found that $R$ varies with the primary energy ($E$) and this provides a
method of estimating the primary energy. In general, one can estimate the
arrival times at each telescope using the radius of curvature, arrival
direction of the primary and the core location.
This, when compared with the data enables us to estimate the above parameters
for each shower. This method of obtaining the arrival direction alleviates
the difficulty in the form of systematics arising out of the plane wavefront
approximation for the \v Cerenkov front. Another outstanding problem in the
field of atmospheric \v Cerenkov technique is the difficulty in locating the
shower core. This method seems to solve both these problems and provides an
elegant method to determine the arrival direction as well as the core location
from timing information alone. In addition, using the \v Cerenkov photon
density information and the core position we can estimate the energy of the
primary if the nature of the primary is known. Combining these two
independent estimates of the primary energy, the energy resolution can be
further improved. Application of this methodology to simulated data and the
results will be presented. The intrinsic uncertainties on the various
estimated parameters also will be discussed.
\end{abstract}

\section{Introduction}

Measurement of shower parameters like the primary energy, core location
and direction of arrival are vital in any air shower experiment. The
density measurements and core location information enable one to estimate the
primary energy which in turn could be used to derive the energy spectrum of
primary. On the other hand, the timing information which enables us to derive
the arrival direction of the primary is essential for source search.

Estimation of shower parameters
is rather easily done at PeV energies by measuring particle densities and their
arrival times. However at TeV energies, where the charged particles do not 
reach the observation level, the measurement of shower parameters is rather a
challenging task. This is primarily because a bulk of the recorded events have
the core locations far away from the points of measurement. In the case of
$\gamma$-ray primaries, the \v Cerenkov photon lateral distribution does not
exhibit significant gradient which makes it even more difficult to estimate the
core location. Further, in the case of \v Cerenkov telescope arrays, in most
cases, only a part of the \v Cerenkov pool is intercepted by the array. As a
result the estimate of the energy of the primary will be very uncertain without
the core location information.

The arrival direction of the primary can be estimated from the timing
measurements of the \v Cerenkov light front at spaced telescopes. The
conventional plane front approximation leads to significant systematic
errors in the estimated angles. These errors often undermine the otherwise
excellent angular resolution of the system. In order to eliminate the
interference due to systematic errors one has to use a spherical approximation
to the light front. As a result, we can obtain important additional
information like the radius of curvature of the \v Cerenkov front and the
core location.

Here we present the methodology and results of such an effort based on
simulation studies.

\section{\v Cerenkov shower front fitting for vertical showers}

In a typical wavefront sampling experiment arrival time of \v
Cerenkov shower front is recorded at several locations in \v Cerenkov
pool with high precision. This information can be used to reconstruct
the shower direction.
It is shown earlier that for vertically incident showers initiated
by $\gamma-$rays and cosmic ray primaries, the relative arrival time delay
[$t(r)$] of \v Cerenkov shower front at a core distance $r$  can be
approximated by
\begin{equation}
 t(r) = {\sqrt {(R^2+r^2)} \over {c}} - {{R} \over {c}} 
\end{equation}
\noindent where $R$ is the radius of curvature of the spherical front
(Battistoni {\it et al.} (1998), Chitnis and Bhat (1999)).

In effect, \v Cerenkov shower front can be approximated with a wavefront
moving at a speed of light, originating from a single point on shower axis.
Figure 1 shows the variation of mean arrival time of \v Cerenkov shower front
as a function of core distance for $\gamma-$rays, protons and Fe nuclei
of various energies incident vertically at the top of the atmosphere. These
showers were simulated using a package known as CORSIKA (Heck {\it et al.}, 
1998).
For each case, arrival time is averaged over typically 10 showers. Best fit
spherical wavefront is also indicated by smooth curve. Radius of the fitted
wavefront ($R$) approximately corresponds to the height of shower maximum
($h_e$) from observation level. This is expected since large fraction of
emission arises from the region in the vicinity of shower maximum.

\begin{figure}[t]
\vspace*{2.0mm} 
\includegraphics[width=8.3cm]{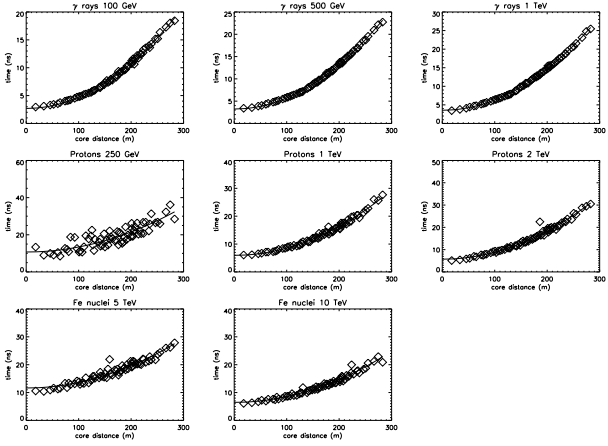}
\caption{Variation of mean arrival time of \v Cerenkov shower front
as a function of core distance for $\gamma-$rays, protons and Fe nuclei
of various energies incident vertically at the top of the atmosphere.
For each case, arrival time is averaged over typically 10 showers. The
smooth curve corresponds to the best fit spherical wavefront.}
\end{figure}

\section{Algorithm for spherical wavefront fitting for inclined showers}

Here we extend this approach to the showers generated by primaries incident
at an angle with respect to the vertical. Consider a distributed array of 
telescopes at observation level. Shower is incident at a
zenith angle $\theta$ and azimuth angle $\phi$. Here it is assumed that
the shower axis passes through the centre of the array. Coordinates of
the telescopes in the array are given by ($x_i$,$y_i$), with respect to
the centre of the array.

The arrival time of the shower front at a telescope situated at
($x_i$,$y_i$), with respect to that at the footprint of the
front, is given by
\begin{equation}
 t(r)={{1} \over {c}} \left[ {\sqrt {{{R^2} \over {cos^2 \theta}} - 2 R tan \theta \left( x_i cos \phi + y_i sin \phi \right) + x_i^2 + y_i^2}} - R \right]
\end{equation}
\noindent where
      $R$ is the vertical height of the emission point on shower axis from the
observation level.

Using this algorithm one can fit the measured arrival time and
estimate the arrival direction
of shower front ($\theta$ and $\phi$) as well as the radius of curvature
of the shower front ($R'$).

\section{Spherical wavefront fitting of simulated data}

Efficacy of this algorithm is tested using simulated data. A large number
of showers initiated by protons incident at an average angle of 15$^\circ$
with respect to vertical are simulated using CORSIKA (Heck {\it et al.}, 
1998).  Energies of these
proton primaries are chosen randomly in the range 250 GeV - 20 TeV with
the spectral index of -2.65. Zenith angles for showers are chosen randomly
within 2$^\circ$ around the average value. Whereas azimuthal angles are
randomly selected in the range 0-360$^\circ$. Axes of all the showers pass
through the central telescope of the array. Arrival times of \v Cerenkov
photons are recorded at 357 telescopes spread over an area of 400 $m$ 
$\times$ 400 $m$ at an altitude of 1 $km$.

For each shower, arrival times at various core distances are fitted
with a spherical wavefront. Angles $\theta$ and $\phi$ and vertical
height of emission point from observation level ($R$) are estimated using
Marquardt routine. This exercise is carried out for a sample of 50 showers.
Figure 2(a) shows the distribution of fitted radii of curvature ($R$) of
shower front. For comparison, distribution of the shower maxima $h_e$
from observation level is also shown in the same figure. Distribution
of difference in $R$ and $h_e$ is shown in Figure 2(b). Mean value of this
difference distribution is 3.3 $km$ with an RMS spread of 2.8 $km$, indicating
that effective point of emission is above the shower maximum. 
Difference distribution of fitted azimuthal angle ($\phi$) and the actual
azimuthal angle is shown in Figure 3a. Similarly difference distribution of
fitted zenith angle ($\theta$) and actual zenith angle is shown in Figure 3b.

\begin{figure}[t]
\vspace*{2.0mm} 
\includegraphics[width=8.3cm]{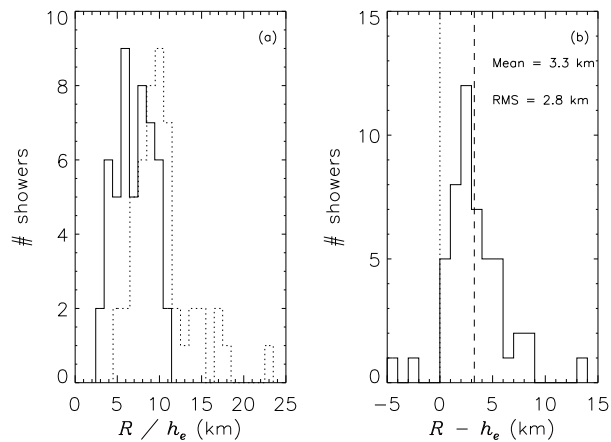}
\caption{(a) Distribution of the fitted radii of shower front $R$ (continuous line)
and distribution of the actual shower maxima $h_e$ from observation level
(shown by dotted line).
(b) Difference distribution of $R$ and $h_e$. Mean of the distribution and
the expected value are marked by dashed and dotted lines. Mean and RMS of
the difference distribution are indicated.}
\end{figure}

So far we have carried out spherical wavefront fitting assuming shower
core to be exactly at the centre of the array, which is also the origin
of our coordinate system. However, in an actual experiment shower core
may lie anywhere in the telescope array and at times even outside the
array. By introducing the core location as another variable, same
algorithm can be further extended to get location of the shower core.

\begin{figure}[t]
\vspace*{2.0mm} 
\includegraphics[width=8.3cm]{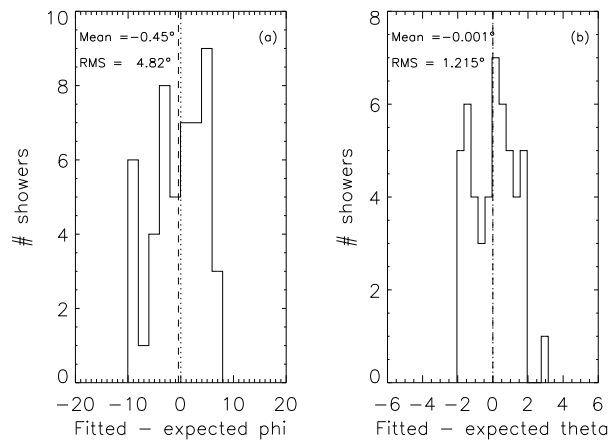}
\caption{(a) Difference distribution of fitted and actual $\phi$. Mean of the
distribution and the expected value are marked by dashed and dotted
lines. Mean and RMS of the difference distribution are indicated.
(b) Difference distribution of fitted and actual $\theta$.  Mean of
the distribution and the expected value are marked by dashed and dotted
lines. Mean and RMS of the difference distribution are indicated.}
\end{figure}

\section{Discussion and conclusions}

\v Cerenkov shower fronts from proton showers incident at an angle
of 15$^\circ$ are fitted with a spherical wavefront. Fitted radius $R$ is
found to be less than the height of shower maximum $h_e$ by about 3 $kms$.
For vertical showers $R$ is found to be about 1 $km$ lower than $h_e$
(Chitnis and Bhat, 2001). Larger difference between $R$ and
$h_e$ for inclined showers could be intrinsic. Reduction in the relative
contribution of the curvature to the arrival time differences at various
telescopes compared to those due to propagation delay could result in a
larger error in fitted radius as well.
One can subtract the relative time of arrival differences arising purely out
of spatial separation of the detectors using plane front approximation. The
residuals can then be fitted to a spherical front which could enhance the
sensitivity of the curvature of the shower front which in turn will reduce
the systematic error on the fitted radius of curvature. We plan to implement
this in our future fitting procedures when we will be able to present a
quantitative estimate of the degree of improvement in the fitted radius of
curvature.

For a given species the height of shower maximum decreases as the logarithm
of the primary energy (Rahman {\it et al.}, 2001). This offers us an alternate
method of estimating the primary energy which in turn improves the primary
energy estimate.

In addition to the above, spherical fit to the \v Cerenkov light front offers
arrival direction as well as core location information of the shower. As
mentioned in the introduction the arrival direction information derived by
this method is relatively free of systematic errors and hence one can
fully exploit the actual angular resolution of the experiment limited
by the system hardware. This, in turn offers a major advantage of rejecting
a bulk of the off-axis showers which could be presumed to be due to cosmic
rays with better confidence.

Finally, shower parameters estimated by this methodology like the radius of
curvature and core location coupled with measured and simulated photon density
distributions will greatly improve the estimation of the primary energy.

\begin{acknowledgements}
We would like to acknowledge the fruitful discussions with Profs.
B. S. Acharya, K. Sivaprasad and P. R. Vishwanath during the present
work.
\end{acknowledgements}

\end{document}